\documentclass[traditabstract,letter,longauth]{aa}
\usepackage{graphicx}
\usepackage{txfonts}
\begin{document}

\title{First results from HerMES on the evolution of the submillimetre
luminosity function\thanks{{\it Herschel} is an ESA
space observatory with science instruments provided by European-led Principal Investigator consortia and with important participation from NASA.}}

\author{S.~Eales\inst{1}
\and G.~Raymond\inst{1}
\and I.\,G.~Roseboom\inst{2}
\and B.~Altieri\inst{3}
\and A.~Amblard\inst{4}
\and V.~Arumugam\inst{5}
\and R.~Auld\inst{1}
\and H.~Aussel\inst{6}
\and T.~Babbedge\inst{7}
\and A.~Blain\inst{8}
\and J.~Bock\inst{8,9}
\and A.~Boselli\inst{10}
\and D.~Brisbin\inst{11}
\and V.~Buat\inst{10}
\and D.~Burgarella\inst{10}
\and N.~Castro-Rodr{\'\i}guez\inst{12,13}
\and A.~Cava\inst{12,13}
\and P.~Chanial\inst{7}
\and D.\,L.~Clements\inst{7}
\and A.~Conley\inst{14}
\and L.~Conversi\inst{3}
\and A.~Cooray\inst{4,8}
\and C.\,D.~Dowell\inst{8,9}
\and E.~Dwek\inst{15}
\and S.~Dye\inst{1}
\and D.~Elbaz\inst{6}
\and D.~Farrah\inst{2}
\and M.~Fox\inst{7}
\and A.~Franceschini\inst{16}
\and W.~Gear\inst{1}
\and J.~Glenn\inst{14}
\and E.\,A.~Gonz\'alez~Solares\inst{17}
\and M.~Griffin\inst{1}
\and M.~Harwit\inst{19}
\and E.~Hatziminaoglou\inst{20}
\and J.~Huang\inst{21}
\and E.~Ibar\inst{22}
\and K.~Isaak\inst{1}
\and R.\,J.~Ivison\inst{22,5}
\and G.~Lagache\inst{23}
\and L.~Levenson\inst{8,9}
\and C.\,J.~Lonsdale\inst{24}
\and N.~Lu\inst{8,25}
\and S.~Madden\inst{6}
\and B.~Maffei\inst{26}
\and G.~Mainetti\inst{16}
\and L.~Marchetti\inst{16}
\and G.\,E.~Morrison\inst{27,28}
\and A.\,M.\,J.~Mortier\inst{7}
\and H.\,T.~Nguyen\inst{9,8}
\and B.~O'Halloran\inst{7}
\and S.\,J.~Oliver\inst{2}
\and A.~Omont\inst{28}
\and F.\,N.~Owen\inst{24}
\and M.\,J.~Page\inst{29}
\and M.~Pannella\inst{24}
\and P.~Panuzzo\inst{6}
\and A.~Papageorgiou\inst{1}
\and C.\,P.~Pearson\inst{30,31}
\and I.~P{\'e}rez-Fournon\inst{12,13}
\and M.~Pohlen\inst{1}
\and J.\,I.~Rawlings\inst{29}
\and D.~Rigopoulou\inst{30,32}
\and D.~Rizzo\inst{7}
\and M.~Rowan-Robinson\inst{7}
\and M.~S\'anchez Portal\inst{3}
\and B.~Schulz\inst{8,25}
\and Douglas~Scott\inst{18}
\and N.~Seymour\inst{29}
\and D.\,L.~Shupe\inst{8,25}
\and A.\,J.~Smith\inst{2}
\and J.\,A.~Stevens\inst{33}
\and V.~Strazzullo\inst{24}
\and M.~Symeonidis\inst{29}
\and M.~Trichas\inst{7}
\and K.\,E.~Tugwell\inst{29}
\and M.~Vaccari\inst{16}
\and I.~Valtchanov\inst{3}
\and L.~Vigroux\inst{28}
\and L.~Wang\inst{2}
\and R.~Ward\inst{2}
\and G.~Wright\inst{22}
\and C.\,K.~Xu\inst{8,25}
\and M.~Zemcov\inst{8,9}}

\institute{Cardiff School of Physics and Astronomy, Cardiff University, Queens Buildings, The Parade, Cardiff CF24 3AA, UK\\
 \email{Steve.Eales@astro.cf.ac.uk}
\and Astronomy Centre, Dept.\ of Physics \& Astronomy, University of Sussex, Brighton BN1 9QH, UK
\and Herschel Science Centre, European Space Astronomy Centre, Villanueva de la Ca\~nada, 28691 Madrid, Spain
\and Dept.\ of Physics \& Astronomy, University of California, Irvine, CA 92697, USA
\and Institute for Astronomy, University of Edinburgh, Royal Observatory, Blackford Hill, Edinburgh EH9 3HJ, UK
\and Laboratoire AIM-Paris-Saclay, CEA/DSM/Irfu - CNRS - Universit\'e Paris Diderot, CE-Saclay, pt courrier 131, F-91191 Gif-sur-Yvette, France
\and Astrophysics Group, Imperial College London, Blackett Laboratory, Prince Consort Road, London SW7 2AZ, UK
\and California Institute of Technology, 1200 E.\ California Blvd., Pasadena, CA 91125, USA
\and Jet Propulsion Laboratory, 4800 Oak Grove Drive, Pasadena, CA 91109, USA
\and Laboratoire d'Astrophysique de Marseille, OAMP, Universit\'e Aix-marseille, CNRS, 38 rue Fr\'ed\'eric Joliot-Curie, 13388 Marseille cedex 13, France
\and Space Science Building, Cornell University, Ithaca, NY, 14853-6801, USA
\and Instituto de Astrof{\'\i}sica de Canarias (IAC), E-38200 La Laguna, Tenerife, Spain
\and Departamento de Astrof{\'\i}sica, Universidad de La Laguna (ULL), E-38205 La Laguna, Tenerife, Spain
\and Dept.\ of Astrophysical and Planetary Sciences, CASA 389-UCB, University of Colorado, Boulder, CO 80309, USA
\and Observational  Cosmology Lab, Code 665, NASA Goddard Space Flight  Center, Greenbelt, MD 20771, USA
\and Dipartimento di Astronomia, Universit\`{a} di Padova, vicolo Osservatorio, 3, 35122 Padova, Italy
\and Institute of Astronomy, University of Cambridge, Madingley Road, Cambridge CB3 0HA, UK
\and Department of Physics \& Astronomy, University of British Columbia, 6224 Agricultural Road, Vancouver, BC V6T~1Z1, Canada
\and 511 H street, SW, Washington, DC 20024-2725, USA
\and ESO, Karl-Schwarzschild-Str.\ 2, 85748 Garching bei M\"unchen, Germany
\and Harvard-Smithsonian Center for Astrophysics, MS65, 60 Garden Street,  Cambridge,  MA02138, USA
\and UK Astronomy Technology Centre, Royal Observatory, Blackford Hill, Edinburgh EH9 3HJ, UK
\and Institut d'Astrophysique Spatiale (IAS), b\^atiment 121, Universit\'e Paris-Sud 11 and CNRS (UMR 8617), 91405 Orsay, France
\and National Radio Astronomy Observatory, P.O.\ Box O, Socorro NM 87801, USA
\and Infrared Processing and Analysis Center, MS 100-22, California Institute of Technology, JPL, Pasadena, CA 91125, USA
\and School of Physics and Astronomy, The University of Manchester, Alan Turing Building, Oxford Road, Manchester M13 9PL, UK
and Institute for Astronomy, University of Hawaii, Honolulu, HI 96822, USA
\and Canada-France-Hawaii Telescope, Kamuela, HI, 96743, USA
\and Institut d'Astrophysique de Paris, UMR 7095, CNRS, UPMC Univ.\ Paris 06, 98bis boulevard Arago, F-75014 Paris, France
\and Mullard Space Science Laboratory, University College London, Holmbury St.\ Mary, Dorking, Surrey RH5 6NT, UK
\and Space Science \& Technology Department, Rutherford Appleton Laboratory, Chilton, Didcot, Oxfordshire OX11 0QX, UK
\and Institute for Space Imaging Science, University of Lethbridge, Lethbridge, Alberta, T1K 3M4, Canada
\and Astrophysics, Oxford University, Keble Road, Oxford OX1 3RH, UK
\and Centre for Astrophysics Research, University of Hertfordshire, College Lane, Hatfield, Hertfordshire AL10 9AB, UK}

\abstract{We 
have carried out two extremely deep surveys with 
SPIRE, one of the two cameras on {\it Herschel}, 
at 250\,$\mu$m, close to the peak of the far-infrared background. We have used the
results to investigate the
evolution of the rest-frame 250-$\mu$m luminosity function out to $z =\rm 2$.
We find evidence for strong evolution out to
$z \simeq\rm 1$ but evidence for at most
weak evolution
beyond this redshift. 
Our results suggest that a significant 
part of the stars and metals in the universe today were formed
at $z \preceq\rm 1.4$ in spiral galaxies.} 

\keywords{Galaxies: evoluton -- Galaxies: formation -- Galaxies: high-redshift --
Submillimetre: galaxies}

\titlerunning{Evolution of the submillimetre luminosity function}
\authorrunning{S.\,A.\ Eales et al.}

\maketitle

\section{Introduction}

The discovery 
that approximately half the energy ever radiated by galaxies
is received on Earth in the far-infrared waveband
(Puget et al.\ 1996; Fixsen et al.\ 1998)
implies that galaxies must show
strong evolution that is hidden from
optical telescopes (Gispert, Lagache \& Puget 2000).
A decade ago, deep surveys with the SCUBA camera on the James Clerk Maxwell Telescope
resolved much of the far-infrared background (FIRB) 
at 850 $\mu$m into individual sources (Barger et al.\ 1998; Hughes
et al.\ 1998). These sources are mostly extremely
luminous dust-enshrouded galaxies at $\rm z > 2$ (Chapman et al.\ 2005) with an average implied star-formation
rate (if the ultimate source of the energy is star formation) of $\simeq$400 $\rm M_{\odot}\ year^{-1}$ (Coppin
et al.\ 2006),
much greater than the star-formation rates in galaxies like our own.

However, the energy density in the FIRB at 850 $\mu$m is
$\simeq$30 times less than at 200 $\mu$m where the FIRB is at a maximum,
and both
the spectral shape of the FIRB 
and statistical `stacking' analyses
(Dole et al.
2006; Pascale et al.\ 2009) imply that much of the FIRB is actually
produced by sources
at lower redshift (Gispert et al.\ 2000; Dole et al.
2006; Pascale et al.\ 2009).
The launch of the Herschel Space Observatory
(Pilbratt et al.\ 2010)
in May 2009 has given us the opportunity
to resolve a significant fraction of the FIRB at wavelengths
where its energy density is at a maximum. In this letter, using the first data
from the 
Herschel Multi-tiered Extragalactic Survey (HerMES; Oliver et al., in prep),
we investigate the evolution implied by the existence of the FIRB by
measuring the evolution of the galaxy luminosity function at 250 $\mu$m.
We everywhere assume a standard concordance cosmology: $\rm \Omega_M=0.28,\ \Omega_{\Lambda}=0.72,\
H_0 = 72\ km\ s^{-1}\ Mpc^{-1}$.

\section{The data}

The images that we analyse in this 
letter were taken at 250 $\mu$m with the SPIRE instrument on {\it Herschel}, whose
in-orbit performance and scientific capabilities are described in
Griffin et al.\ (2010). The calibration methods and accuracy of SPIRE are
described by Swinyward et al.\ (2010).
The three images
consist
of a shallow image of the Lockman Hole (LH) and deep images of the northern
field of the Great Observatories Origins Deep Survey (GOODS-North) and of a field
within the Lockman Hole (LH-North). For the latter two images
the dominant source of noise is confusion due to numerous faint
sources, which is 5.8 mJy beam$^{-1}$ at 250 $\mu$m (Nguyen et al.\ 2010).

\begin{figure}
   \centering
   \includegraphics[width=6cm]{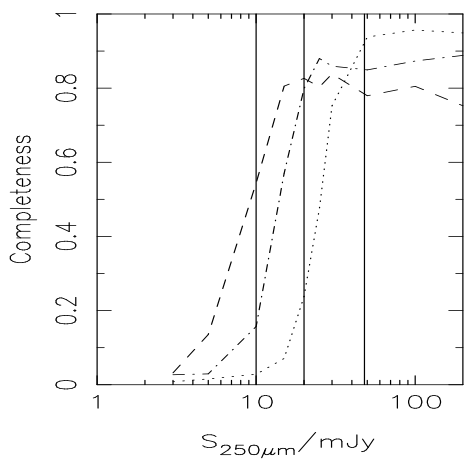}
      \caption{Fraction of sources recovered by the source-extraction
method as a function of flux density for GOODS-North (dashed), 
LH-North (dot-dashed) and the LH (dotted). The vertical lines
show the flux limits we have used for the samples from the three
fields (Table~1).}
   \end{figure}

\begin{table}
\caption{Field Statistics}             
\label{table:1}      
\centering                          
\begin{tabular}{l c c c}        
\hline\hline                 
Field & GOODS-North & LH-North & LH \\    
\hline                        
Area (deg$^2$) & 0.031 & 0.371 & 5.17 \\      
Flux limit (mJy) & 10 & 20    & 48 \\
No.\ sources & 83 & 276 & 551 \\
Spectroscopic z & 65 & 75 & 140 \\
Photometric z & 18 & 161 & 322 \\ 
No z (\%) & 0 & 14 & 16 \\

\hline                                   
\end{tabular}
\end{table}

To measure robust fluxes for sources close to the confusion level, we
have developed a source-extraction 
technique that is based on the assumption that sources detected
at 250 $\mu$m with SPIRE will also be detected in deep observations with {\it Spitzer} at
24 $\mu$m. This assumption was suggested by the recent studies that concluded
that galaxies detected  in deep 24-$\mu$m surveys with {\it Spitzer}
produce most of the FIRB at 160 $\mu$m (Dole et al.\ 2006) and
at 250, 350 and 500 $\mu$m (Marsden et al.\ 2009). 
This approach reduces the effective confusion noise by resolving some of the
confusing background into the sources detected at 24 $\mu$m.
Our 
`cross-ID method' starts
from a list of the 24-$\mu$m sources found in the field covered by the 250-$\mu$m image. By using
a matrix inversion technique, we then find 
the 250-$\mu$m fluxes at these positions that provide the best fit to the 250-$\mu$m image (Roseboom,
in prep). A problem with methods like this is that there can be a large number of degenerate
solutions when the surface-density of sources in the input catalogue is large, as it is with the
deep {\it Spitzer} 24-$\mu$m catalogues. We have addressed this problem by iteratively reducing the number
of sources in the input catalogue in order to find the list of input sources that provides the best fit
to the 250-$\mu$m image.
Full details of the method, its validation with simulations, and a comparison of the results with
other methods of source extraction are given in Roseboom et al.\ (in prep).

\begin{figure}
   \centering
   \includegraphics[width=6cm]{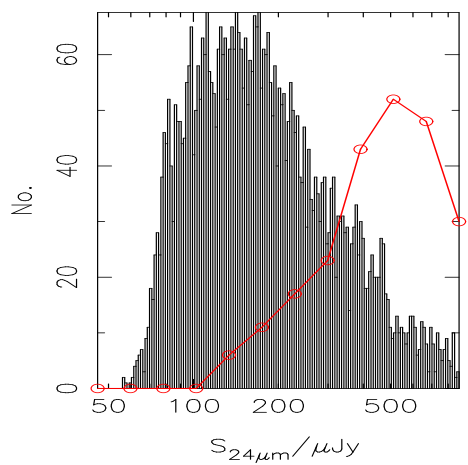}
      \caption{The histogram
shows the distribution of 24-$\mu$m flux density for the
5015 24-$\mu$m sources
in the {\it Spitzer} catalogue
used in the source extraction for LH-North. 
The red line shows the 
distribution of 24-$\mu$m flux density for the sample of 276 250-$\mu$m sources 
found using our source-extraction method.}
   \end{figure}

Roseboom et al.\ (in prep) have assessed the completeness of the 
cross-ID method by inserting artificial sources onto the real images and then determining the
fraction of these sources that are detected by the source-extraction method. Fig.~1
shows the results for the three fields as a function
of flux density. Using these curves to choose the
flux limits, we have used the cross-ID method to extract
samples of sources from 
regions within each field for which the optical/IR data is
particularly good (Table 1). We accepted a lower completeness level at the
flux limit for GOODS-North ($\simeq$50\%) in order to get the maximum range of
luminosity at each redshift. When calculating the luminosity function (\S 3),
we used the curves in Fig.~1 to correct for incompleteness. 
All the sources above these flux limits are detected at $\bf \ge 5 \sigma$.

Fig.~1 does not account for any objects that are missing because
their 24-$\mu$m flux density
falls below the limit of the {\it Spitzer} catalogue used in the cross-ID method. 
There are
two arguments that suggest this is not a serious problem. First, Roseboom et al.\ 
show that the 
number-density of 250-$\mu$m sources in the cross-ID catalogues agree well with the
source counts determined by Oliver et al.\ (2010) down to $\simeq$25 mJy in
LH-North and $\simeq$40 mJy in LH.
Second, they 
compare the results of using a shallow and
a deep {\it Spitzer} catalogue in several fields and conclude that using the deeper catalogue does
not produce a large increase in the number of 250-$\mu$m sources 
found by the cross-ID method. Fig.~2 illustrates this 
clearly for the LH-North field.
The figure shows a histogram of the 24-$\mu$m flux densities of the
sources in the input {\it Spitzer} catalogue, 
whereas the red line shows the 24-$\mu$m
flux densities of the objects found by the cross-ID method.
It is important to note that the 24-$\mu$m flux densities of the
sources in the input catalogue are not used as information in the
cross-ID method, so the very different distributions for the input
catalogue and for the objects found by the cross-ID method is evidence
that the method is working well. The small number of objects with faint
24-$\mu$m flux densities found by the cross-ID method, despite the
very large number of faint 24-$\mu$m sources in the input catalogue,
strongly suggests that we are not missing objects that have too low
24-$\mu$m flux densities to be included in the {\it Spitzer} catalogues.

\begin{figure}
   \centering
   \includegraphics[width=6cm]{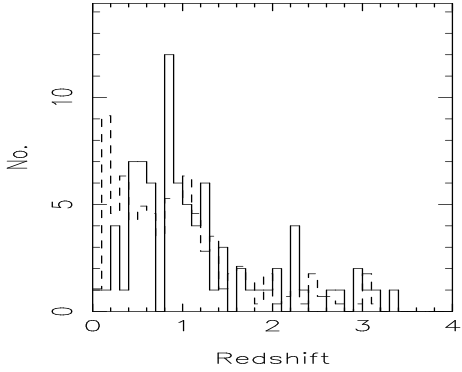}
      \caption{Redshift histograms for GOODS-North (continuous) and LH-North
(dashed). The values in the histogram for LH-North have been scaled to the
number of sources found in GOODS-North.}
   \end{figure}

We have used the optical and near-infrared images
for these fields to find the counterparts to the 24-$\mu$m sources.
Table 1 shows we have redshifts for all of the GOODS-North sources, either a spectroscopic redshift from the
catalogue of Barger et al.\ (2008) or, for 22\% of the sources, a photometric 
redshift, which we have estimated
from the available 
multi-wavelength images
(in typically nine optical and near-infrared bands -- Raymond et al., in prep).
The situation 
for our other deep field is a little less satisfactory, although we
still have redshifts for 86\% of the sources, a mixture 
of spectroscopic redshifts and photometric redshifts, mostly taken
from an unpublished catalogue of Owen and collaborators, which was
produced from images
in eight optical and near-infrared bands. 
The data for LH is described in Rowan-Robinson et al.\ (2008) and Vaccari
et al.\ (in prep).
Fig.~3 shows the redshift distributions for our two deep samples.
The differences between the two can probably be explained by cosmic variance
as the result of the small area of the GOODS-North sample. The similarity
of the distributions at high redshift suggests
that the sources without redshifts in LH-North are not preferentially 
at high redshift.

\begin{figure}
   \centering
   \includegraphics[width=9cm]{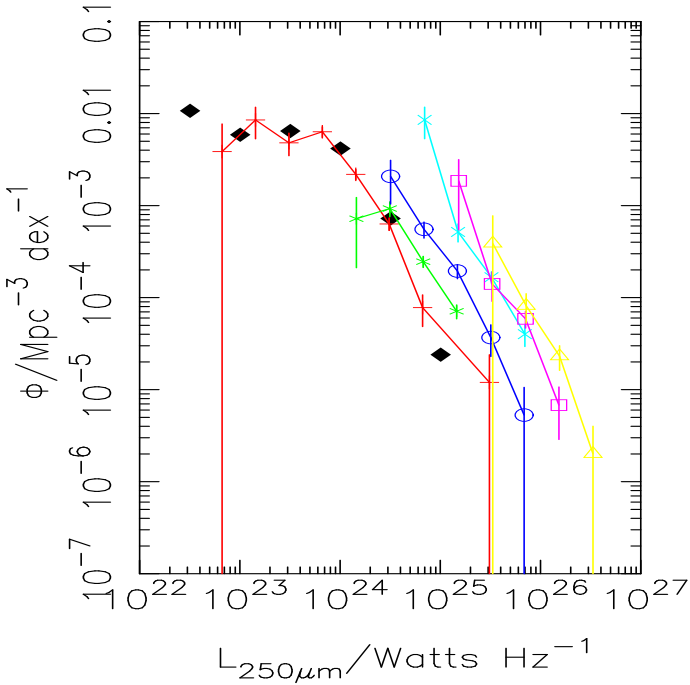}
      \caption{Luminosity function at 250 $\mu$m in six redshift
slices: $\rm 0.0 < z \le 0.2$---red; $\rm 0.2 < z \le 0.4$---green; $\rm 0.4 < z \le 0.8$---dark
blue; $\rm 0.8 < z \le 1.2$---light blue; $\rm 1.2 < z \le 1.6$---purple; $\rm 1.6 < z \le 2.0$---yellow.
The black diamonds show the estimate of the local 250-$\mu$m luminosity function from
Vaccari et al.\ (2010).}
   \end{figure}

\section{Estimates of the luminosity function}

We have estimated the rest-frame 250-$\mu$m luminosity function in six redshift intervals:
$\rm 0.0 < z \le 0.2$, $\rm 0.2 < z \le 0.4$, $\rm 0.4 < z \le  0.8$, $\rm 0.8 < z \le 1.2$, 
$\rm 1.2 < z \le 1.6$, $\rm 1.6 < z \le 2.0$. 
Because the optical/IR data is less sensitive for this
field than the other two, we have only used the LH sample for the two
low-redshift intervals.
We have not estimated the luminosity function at $\rm z > 2$ because of the low
number of sources above this redshift.

At this early stage of the HerMES project, we do not
have large numbers of measurements of 
the spectral energy distributions (SED) of individual galaxies.
We have therefore
made the assumption that all galaxies have the same SED,
which we have taken to be a grey body with a dust-emissivity index of 1.5 and a temperature of 26 K, the average of the 
SEDs found for the galaxies detected at the same wavelength 
by the Balloon-borne Large-Aperture Sub-millimeter Telescope (BLAST; Dye et al.\ 2009). 
If the typical temperature of the dust in a SPIRE
galaxy does change systematically with
redshift, this will 
have little effect on the shape of a 
galaxy SED at $\rm \lambda > 100\ \mu m$ and
thus on the luminosity functions at $\rm z < 1$ but it might have a significant 
effect on the luminosity functions at higher redshift.

We have estimated the luminosity function using the estimator of
Page and Carrera (2000), which is not only the ideal estimator in general but 
is particularly well-suited to situations where
source confusion is an issue (Eales et al.\ 2009). 
For each source, we used the curves in Fig.~1 to 
estimate a correction factor, $C_i$, which is the reciprocal of the
completeness at the flux density of the source.
Our estimate of the value of the
luminosity function
in a bin of the luminosity-redshift plane is then $\sum C_i / V$, where 
the sum is over the galaxies in this bin and $V$ is the accessible volume
in this bin averaged over the width of the bin in luminosity. We have 
combined the data for the different HerMES fields by adopting the
`coherent volume' approach of Avni and Bahcall (1980).
Our estimate of the 
luminosity function in a bin is then $\sum C_i / \sum V$, where 
the sum in the numerator is now over all galaxies in the three fields
in that bin and the sum in the denominator is the sum of the accessible
volumes for the three fields.
The fractional error
on the luminosity function is then $\sqrt{N} / N$, in which $N$ is
the uncorrected number of galaxies 
in that bin in the three fields.
This error does not include the effect of cosmic variance, which
we will consider in a later paper.

Fig.~4 shows our estimates of the luminosity function in the six redshift intervals
and the HerMES estimate of the local luminosity function 
(Vaccari et al.\ 2010), which was estimated over
two
fields covering 14.7 deg$^2$ (including our LH area) and agrees reassuringly
with our estimate 
in the lowest redshift slice.
The luminosity
function shows steady evolution out to $\rm z = 1$, in the sense that the space-density of the most luminous
sources gradually increases with redshift. This agrees 
well with the evolution in the luminosity
function determined from the BLAST results (Eales et al.\ 2009). There 
is evidence in Fig.~4 for 
at most weak evolution at $\rm z > 1$.

There are a number of possible systematic errors
that could affect our results.
One of the nice properties of the cross-ID method is that
simulations suggest it is relatively immune to the effect of flux boosting
due to source confusion and Eddington bias (Roseboom et al., in prep).
A second possible problem is
that we do not have redshifts for all the sources in LH and LH-North.
If the incomplete redshift information affects each redshift bin equally, 
the luminosity functions
will all be slightly too low. However, if the 
incompleteness preferentially affects the high-redshift bins,
we will have under-estimated
the size of the evolution. 
Another potential problem is the large number of photometric
redshifts in our analysis.
We have used a Monte-Carlo simulation to generate artificial samples based on
a no-evolution assumption, giving each source a photometric redshift
with an accuracy equal to our least accurate redshifts, 
and then used the artificial samples to estimate
the luminosity function in the six redshift
slices. This analysis confirms that
the large number of photometric redshifts does
not generate
spurious strong evolution. 
Finally, even if the dust in galaxies does not get systematically hotter with redshift,
there will be a selection effect in our results, since at high redshift we are sampling a lower
rest-frame wavelength, and the galaxies found in samples selected at shorter wavelengths tend
to contain hotter dust. The effect of this selection effect is that we may have overestimated slightly
the luminosity functions at $\rm z > 1$.  

\section{Discussion}  

On the assumption that the star-formation rate is proportional to the
rest-frame 250-$\mu$m luminosity, the lack of evidence for any strong
evolution at $\rm z > 1$ is consistent with investigations that have concluded that
the overall star-formation rate per unit comoving volume was approximately
constant at $\rm z > 1$
(Steidel et al.\ 1999; Gispert et al.\ 2000; Hopkins 2004).
We estimate from the GOODS-North catalogue that we have resolved $\simeq$20\% of the
FIRB at 250 $\mu$m. Fig.~3 shows that the sources making up this top 20\% of the FIRB
are at moderate redshift ($z \sim 1$). Although we cannot say anything directly about
the sources responsible for the missing 80\% of the FIRB, the stacking analysis
of Pascale et al.\ (2009) from BLAST data suggests
that the remainder of the FIRB at this wavelength is also from sources
at moderate redshift.
 
A revealing way to look at these early HerMES results is to use the relationship
between the production of metals and the background radiation associated with
this metal production (Peacock 1993). Using a value for the integrated FIRB of
$\rm 14\ nW\ sr^{-1}$ (Fixsen et al.\ 1998), we obtain a 
relationship between the mass of metals
and the fraction, $\epsilon$, of the FIRB produced in a particular redshift
interval:
\smallskip
$$
M_{metals} = 9 \times 10^6 \epsilon (1+z)\ M_{\odot}\ Mpc^{-3}.
$$
\smallskip
\noindent Fig.~5 shows the metals produced as a function of redshift on the
assumption that the redshift distributions of GOODS-North and LH-North are representative
of the FIRB as a whole. The figure suggests that most of the metals (and therefore most of the
massive stars) formed at a moderate redshift ($z \preceq 1.4$). 
This conclusion is consistent with the overall star-formation rate in the universe
being constant at $\rm z > 1$, and the apparent
decline of metal production at high redshift
is simply the consequence of the relationship between cosmic time and redshift.
However, the decline in the figure at high redshifts 
is relatively small, and since we are
making the large assumptions
that the redshift distributions of our samples
are characteristic of the FIRB as a whole, it is possible that observations that
resolve more of the FIRB and at several
wavelengths rather than a single wavelength will modify this conclusion.

The excellent images that exist from the Hubble Space Telescope for the GOODS-North   
{\it Herschel} sources mean that we can examine the nature of the galaxies responsible for
this metal production.
We have used the z-band images, since at $\rm z \simeq 1$ this band corresponds in
the rest-frame to the B-band. Of the 26 galaxies in the redshift interval $\rm 0.8 < z < 1.2$, 
12 show clear signs of spiral structure, and there may well be more if some of the galaxies
have spiral arms that are below the surface-brightness detection threshold of the images.
Thus the galaxies found in the first HerMES
images appear to be quite different from the major mergers found in the
SCUBA surveys (Ivison et al.\ 2000; Tacconi et al.\ 2008).
Studies of galaxy evolution based on {\it Spitzer} 24-$\mu$m surveys
have also found that the {\it Spitzer} galaxies at $\rm z \simeq 1$ are
often spirals rather than major
mergers (Elbaz et al.\ 2007) but the crucial
advance that {\it Herschel} has made possible
is that we can determine the cosmological importance of these objects, in terms
of the total masses of stars and metals formed in them.

\begin{figure}
   \centering
   \includegraphics[width=6cm]{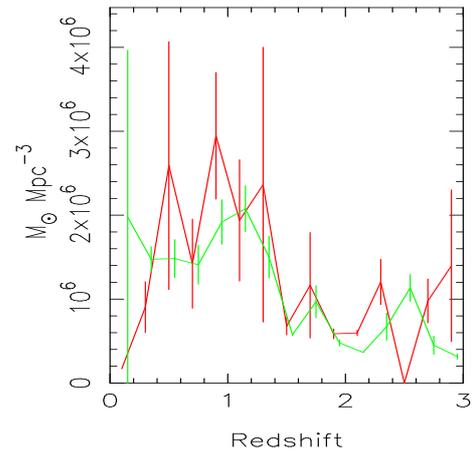}
      \caption{An estimate of the total metals formed per cubic Mpc in different
redshift intervals using the method described in the text. We have made the
estimates from the redshift distributions of GOODS-North (red) and LH-North
(green) on the assumption that these redshift distributions represent the
FIRB as a whole.}
   \end{figure}

\acknowledgements{
SPIRE has been developed by a consortium of institutes led by
Cardiff Univ.\ (UK) and including Univ.\ Lethbridge (Canada);
NAOC (China); CEA, LAM (France); IFSI, Univ.\ Padua (Italy);
IAC (Spain); Stockholm Observatory (Sweden); Imperial College
London, RAL, UCL-MSSL, UKATC, Univ.\ Sussex (UK); Caltech, JPL,
NHSC, Univ.\ Colorado (USA). This development has been supported
by national funding agencies: CSA (Canada); NAOC (China); CEA,
CNES, CNRS (France); ASI (Italy); MCINN (Spain); Stockholm
Observatory (Sweden); STFC (UK); and NASA (USA).}

\end{document}